\documentclass[twocolumn,english]{article}
\usepackage[T1]{fontenc}
\usepackage[utf8]{inputenc}
\usepackage[a4paper]{geometry}
\usepackage{babel}
\usepackage{bm}
\usepackage{amsmath}
\usepackage{amssymb}
\usepackage{graphicx}
\usepackage[unicode=true]
 {hyperref}

\makeatletter
\usepackage{babel}
\usepackage{xcolor}% specify colors by name (e.g. "red")
\definecolor{myred}{rgb}{0.66, 0.15, 0.15}
\definecolor{darkgreen}{rgb}{0.0, 0.5, 0.0}

\usepackage{hyperref}          % clickable URls and cross-references
\hypersetup{
     colorlinks   = true,
     citecolor    = myred,
     urlcolor     = myred,
     urlbordercolor = myred
}

\usepackage[normalem]{ulem}% used to provide various types of underlining
\usepackage{flushend}
\usepackage{cuted}

\usepackage{cite}

\makeatother

\begin{document}
\begin{strip}

{\Large{}\vspace{-1cm}
}\textsf{\textbf{\huge{}Inequivalence of stochastic and Bohmian arrival
times in time-of-flight experiments\vspace{0.5cm}
}}{\huge\par}

{\Large{}Pascal Naidon}\textsf{\textbf{\huge{}\vspace{0.2cm}
}}{\huge\par}

{\small{}Few-Body Systems Physics Laboratory, \href{https://ror.org/05tqx4s13}{RIKEN Nishina Centre},
RIKEN, Wak{ō}, 351-0198 Japan.}\textit{ }{\Large{}\vspace{-0.4cm}
}{\Large\par}

\textit{\href{mailto:pascal@riken.jp}{pascal@riken.jp}}\textsf{\textbf{\huge{}\vspace{0.2cm}
}}{\huge\par}

\today{\Large{}\vspace{0.5cm}
}{\Large\par}

Motivated by a recent prediction~{[}\href{https://doi.org/10.1038/s42005-023-01315-9}{Com. Phys., 6, 195 (2023)}{]}
that time-of-flight experiments with ultracold atoms could test different
interpretations of quantum mechanics, this work investigates the arrival
times predicted by the stochastic interpretation, whereby quantum
particles follow definite but non-deterministic and non-differentiable
trajectories. The distribution of arrival times is obtained from a
Fokker-Planck equation, and confirmed by direct simulation of trajectories.
 It is found to be in general different from the distribution predicted
by the Bohmian interpretation, in which quantum particles follow definite
deterministic and differentiable trajectories. This result suggests
that trajectory-based interpretations of quantum mechanics could be
experimentally discriminated.

\end{strip}

\section{Introduction\label{sec:Introduction}}

Do quantum particles follow definite trajectories? In the textbook
presentations of quantum mechanics~\cite{Landau1989,Tannoudji1973,phillips2003,Shankar2011}
following the standard Copenhagen interpretation of the theory or
its statistical interpretation~\cite{Ballentine1998,Aharonov1961},
emphasis is put on measurements as the only accessible elements of
reality. Classical concepts such as trajectories are thus considered
unnecessary or even inconsistent with observations. In this conception,
a particle is only described by its quantum state in the absence of
any observation, and ``materialises'' only at a certain position
where it is observed by an experimental apparatus at a certain time.

However, it was shown long ago, prominently by Bohm~\cite{Bohm1952,Bohm1952a},
that the formalism of quantum theory is not inconsistent with the
particles having definite trajectories in-between measurements. This
has led to the \emph{de~Broglie-Bohm theory} or \emph{pilot wave
theory}, which  has been recognised as an alternative interpretation
of quantum mechanics~\cite{Bell1987}, in so far as it yields the
same predictions as the standard interpretation. The Bohmian interpretation,
however, has not gained much popularity, notably because it posits
the seemingly unnecessary existence of hidden variables (the particles'
positions at all times), and mainly because, like most interpretations
of quantum mechanics, it does not seem to provide any new testable
prediction.

\begin{figure*}[t]
\includegraphics[height=6cm]{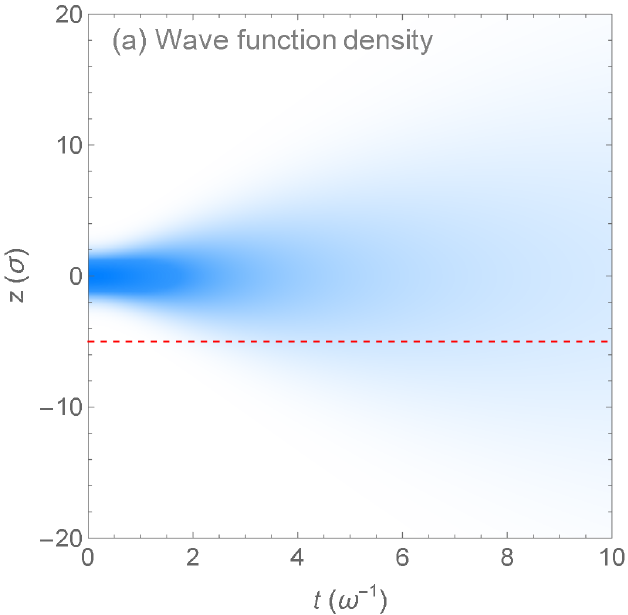}~\includegraphics[viewport=20bp 0bp 301bp 314bp,clip,height=6cm]{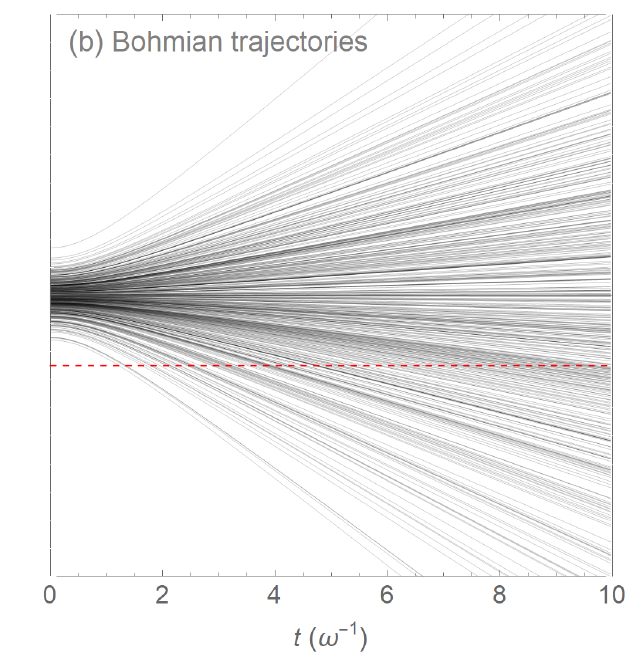}~\includegraphics[viewport=20bp 0bp 301bp 314bp,clip,height=6cm]{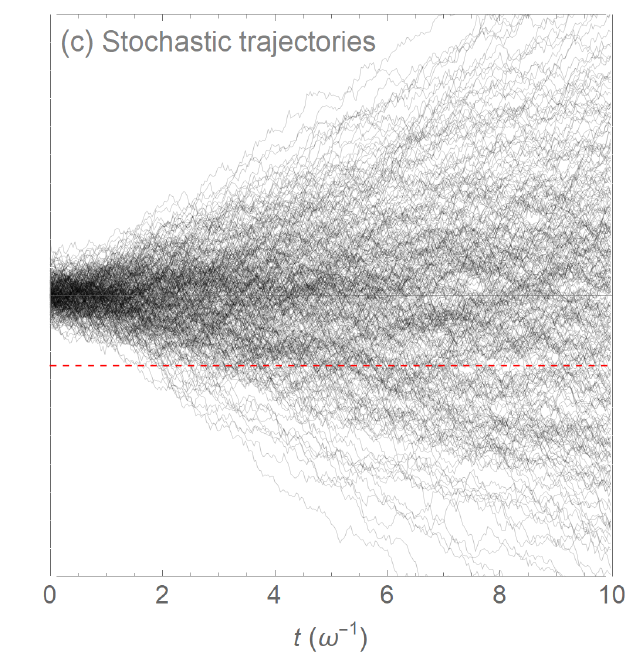}

\caption{\label{fig:Single-Well}Cloud of particles released from a single
harmonic well of frequency $\omega=\hbar/2m\sigma^{2}$ and oscillator
length $\sigma$. (a): integrated density $\int dxdy\vert\psi(\bm{x},t)\vert^{2}$
as a function of time. (b): $z$-component of 400 Bohmian trajectories.
(c): $z$-component of 400 stochastic trajectories. The trajectories
are calculated from Eqs.~(\ref{eq:Bohmian-Trajectory}) and (\ref{eq:Stochastic-Trajectory})
with $dt=1/(1600\omega)$. The red dashed line indicates the position
of a detector at the distance $L=5\sigma$ from the centre of the
well.}
\end{figure*}

\begin{figure*}
\includegraphics[width=6.1cm,height=5.9cm]{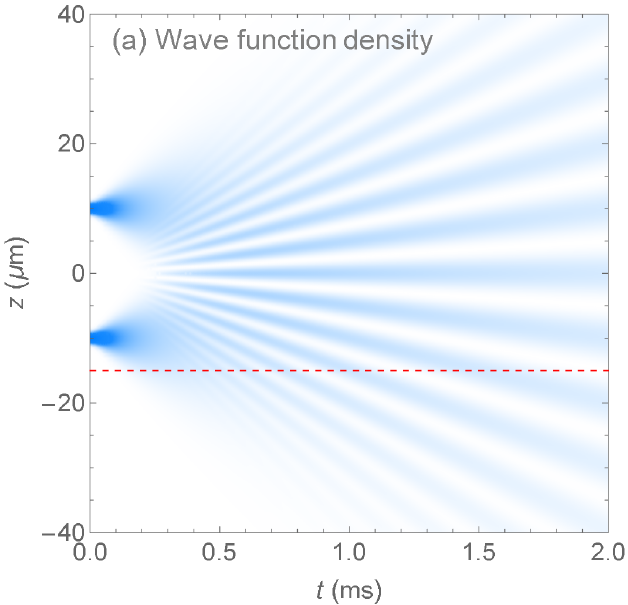}\includegraphics[viewport=10bp 0bp 301bp 310bp,clip,height=5.9cm]{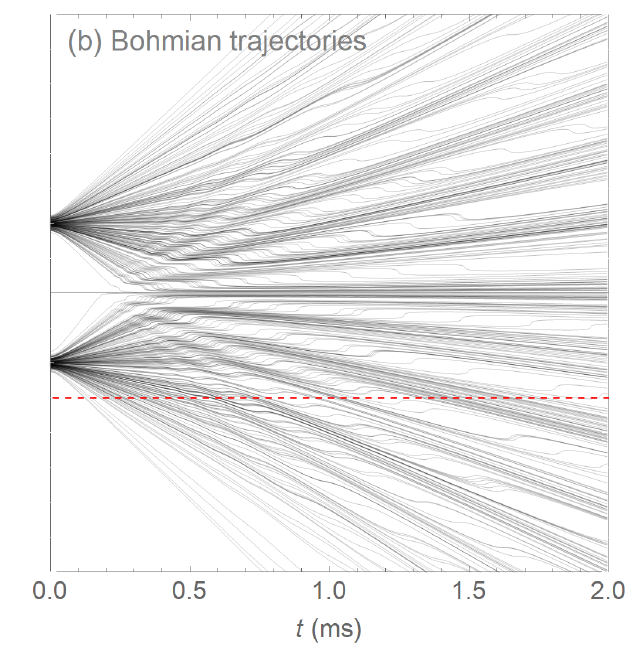}\includegraphics[viewport=15bp 0bp 341bp 351bp,clip,height=5.9cm]{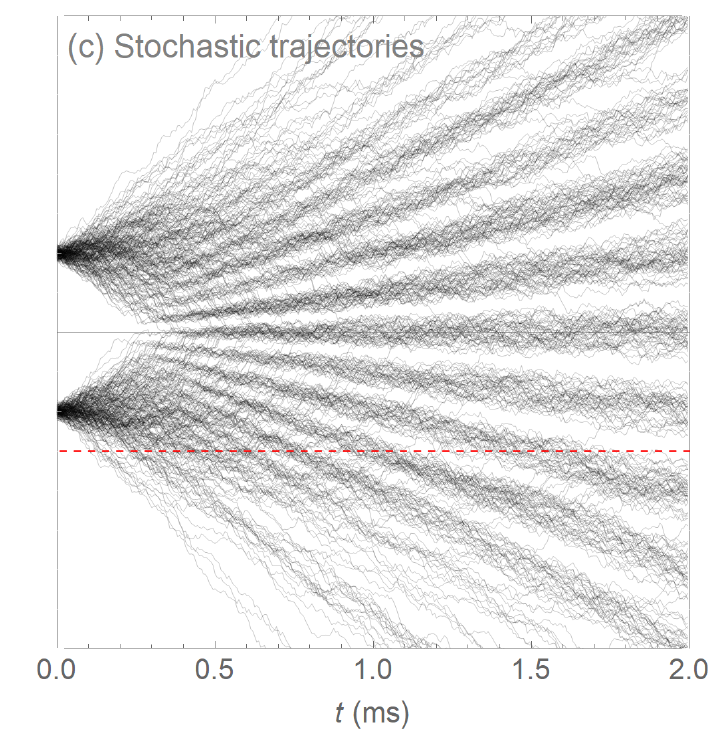}

\caption{\label{fig:Double-Well}Cloud of particles of mass $m=4m_{u}$ released
from two wells of oscillator length $\sigma=0.5\,\mu$m, separated
by a distance $2d=20\,\mu$m along the $z$ direction. The panels
are similar to those of Fig.~\ref{fig:Single-Well}. The red dashed
line indicates the position of a detector at the distance $L=15\,\mu$m
from the centre of the two wells.}
\end{figure*}
It has long been known, however, that there are measurements for which
the interpretations of quantum mechanics may lead to different predictions,
namely, the measurements of arrival times~\cite{Muga2000}. While
the quantum formalism gives the probabilities of a particle's position
measurements at a given time through the square modulus $\vert\psi(\bm{x},t)\vert^{2}$
of its wave function, it does not provide explicit probabilities for
the time $t$ at which a particle arrives at a certain point $\bm{x}$.
One could naïvely think that this probability distribution is still
provided by the square modulus $\vert\psi(\bm{x},t)\vert^{2}$ at
a fixed point $\bm{x}$, but a quick dimensional analysis shows that
it cannot be so: $\vert\psi(\bm{x},t)\vert^{2}$ has the units of
density, whereas the sought probability distribution should be a number
per units of surface and time. The mathematical reason behind this
difficulty is that time is only a parameter in standard quantum mechanics,
whereas the conventional formalism requires measurable quantities
to be described by a self-adjoint operator, and it is known that a
self-adjoint operator cannot be constructed for time~\cite{Pauli1958,Paul1962}.
This problem has led to various efforts to find a plausible way to
predict arrival times in quantum mechanics. Some of these works~\cite{Page1983,Aharonov1984,Rovelli1996,Reisenberger2002,Brunetti2010,Giovannetti2015,Maccone2020}
either extend or reformulate the original formalism of quantum theory
to obtain predictions, while others~\cite{Aharonov1961,Allcock1969a,Allcock1969b,Allcock1969c,Kijowski1974,Werner1986,Grot1996,Delgado1997,Aharonov1998,Baute2000,Galapon2005,Juric2022}
have attempted to obtain predictions within the conventional framework
of quantum theory (although this has been disputed~\cite{Kijowski1999,Leavens2002,Egusquiza2003,Leavens2005}).
On the other hand, in a trajectory-based interpretation of quantum
mechanics such as the Bohmian interpretation, there is seemingly no
difficulty to predict arrival times since the particle is assumed
to follow a definite trajectory, with a definite arrival time at a
certain point~\cite{McKinnon1995,Leavens1998}. As a result, rather
than a mere interpretation it becomes a falsifiable theory in its
own right when applied to the arrival time problem. By measuring arrival
times in time-of-flight experiments, it is therefore possible in principle
to test the different formulations, extensions, and trajectory-based
interpretations of quantum mechanics~\cite{Roncallo2023}.

However, up to now, time-of-flight measurements have only been performed
far from the particle's source of emission, in a regime where all
theories give the same predictions, consistent with a classical motion
of the particle near the detector. The situation may change, however,
as a recent proposal~\cite{AyatollahRafsanjani2023} shows that it
may be possible to discriminate these theories by measuring the arrival
time distribution in a double-slit (or double-well) experiment.

In this context, it is of interest to revisit a rather little known
trajectory-based interpretation of quantum mechanics called \emph{stochastic
mechanics}. Stochastic mechanics started with the realisation by Fényes~\cite{Fenyes1952}
and then Nelson~\cite{Nelson1966} that the Schrödinger equation
naturally appears when considering a certain kind of frictionless
Brownian motion. This led to an attempt to reconstruct quantum theory
from the stochastic motion of particles induced by a hypothetical
fluctuating ether~\cite{Nelson1985}. Reference~\cite{Derakhshani2022}
gives a good account of the current status of stochastic mechanics.
Although the original aim of deriving quantum theory from a more fundamental
theory has not been achieved by stochastic mechanics, it allows for
a given wave function to assign definite (but non-deterministic and
non-differentiable) trajectories to the corresponding particles in
accordance with the predictions of quantum mechanics. From this perspective,
it can be used as an alternative pilot wave theory. This theory
may be regarded as a stochastic version of the de~Broglie-Bohm pilot
wave theory, and we shall call it the \emph{stochastic pilot wave
theory}, to distinguish it from the original stochastic mechanics.

Although the Bohmian and stochastic pilot wave theories are similar,
there appear to have been no detailed comparisons between stochastic
and Bohmian trajectories' arrival times. Previous results~\cite{Aoki2000,Nitta2008}
suggest that stochastic trajectories and Bohmian trajectories lead
to the same arrival time distribution. In this work, it is shown that
they do in fact lead to different arrival time distributions, most
notably in the case of the double-well experiment proposed in Ref.~\cite{AyatollahRafsanjani2023}.
This opens the possibility to evidence, and even characterise, the
trajectories of particles underlying the standard quantum theory.
However, this requires the arrival times of such trajectories to be
faithfully reported by a detecting apparatus, without any substantial
error or perturbation from the detection scheme. The last section
of this article discusses possible issues with actual measurements
of these arrival times.

\section{Bohmian and stochastic trajectories\label{sec:Trajectories}}

The definitions of Bohmian and stochastic trajectories for a given
wave function are closely related. Consider for simplicity, the case
of a single non-relativistic particle of mass $m$, described by a
wave function $\psi(\bm{x},t)$. One can define from the wave function
the complex velocity $\bm{\mathcal{V}}=\frac{\hbar}{m}\bm{\nabla}\ln\psi$
with real part $\bm{u}$ and imaginary part $\bm{v}$ called respectively
\emph{osmotic} and \emph{average} velocities~\cite{Nelson1966}.
Accordingly, one obtains the two probability currents $\bm{i}=\rho\bm{u}$
and $\bm{j}=\rho\bm{v}$, where $\rho$ is the probability density
$|\psi|^{2}$. Note that $\bm{j}$ is the usual probability current
$\frac{\hbar}{m}\text{Im}\left(\psi^{*}\bm{\nabla}\psi\right)$ satisfying
the continuity equation:
\begin{equation}
\frac{\partial\rho}{\partial t}+\bm{\nabla}\cdot\bm{j}=0.\label{eq:Continuity-Equation}
\end{equation}
One may also define the forward and backward drifts,
\begin{align}
\bm{b} & =\bm{v}+\bm{u}\label{eq:Forward-drift}\\
\bm{b}_{*} & =\bm{v}-\bm{u}\label{eq:Backward-drift}
\end{align}
and the corresponding forward and backward currents,
\begin{align*}
\bm{\mathcal{J}} & =\rho\bm{b}\\
\bm{\mathcal{J}}_{*} & =\rho\bm{b}_{*}
\end{align*}

The Bohmian trajectory starting from a point $\bm{x}_{0}$ at time
$t_{0}$ is simply the trajectory that remains tangent to the average
velocity field $\bm{v}$. Namely, the position $\bm{x}^{\prime}$
at time $t^{\prime}=t+dt$ is obtained from the position $\bm{x}$
at time $t$ by the relation:
\begin{equation}
\bm{x^{\prime}}=\bm{x}+\bm{v}(\bm{x},t)dt\label{eq:Bohmian-Trajectory}
\end{equation}
Note that the trajectory is by construction differentiable, uniquely
defined by the starting point, and never intersects any other trajectory
starting from a different point at the same time.

On the other hand, a stochastic trajectory starting from a point $\bm{x}_{0}$
at time $t_{0}$ is defined as a stochastic diffusive process drifting
along the forward velocity field $\bm{b}$. Namely, the position $\bm{x}^{\prime}$
at time $t^{\prime}=t+dt$ is obtained from the position $\bm{x}$
at time $t$ by the relation:
\begin{equation}
\bm{x^{\prime}}=\bm{x}+\bm{b}(\bm{x},t)dt+\bm{\xi}\label{eq:Stochastic-Trajectory}
\end{equation}
where $\bm{\xi}$ is a random vector with average zero and variance
$\frac{\hbar}{m}dt$. Note that, in this case, the trajectories are
non-deterministic, non-differentiable, and may intersect.

It has be shown for both Bohmian~\cite{Bohm1953} and stochastic~\cite{Nelson1966,Nelson1985}
trajectories that when starting from an ensemble of points $\bm{x}_{0}$
at time $t_{0}$ distributed according to the initial density distribution
$\rho(\bm{x},t_{0})$, the subsequent positions along the trajectories
at a later time $t$ are distributed according to the density $\rho(\bm{x},t)$,
in accordance with the predictions of standard quantum mechanics.
This is illustrated in Figs.~\ref{fig:Single-Well} and \ref{fig:Double-Well}
for the case of an ensemble of particles initially confined in the
ground state of a single harmonic well of oscillator lengths $\sigma_{x},\sigma_{y},\sigma_{z}\equiv\sigma$,
with a wave function given by:

\begin{equation}
\psi(\bm{x},t)=G_{\sigma_{x}}(x,t)G_{\sigma_{y}}(y,t)G_{\sigma}(z,t)\label{eq:Single-Well-Wavefunction}
\end{equation}
and in the ground state of two degenerate harmonic wells separated
by distance $2d$ in the vertical direction $z$, with wave function:

\begin{equation}
\psi(\bm{x},t)=G_{\sigma_{x}}(x,t)G_{\sigma_{y}}(y,t)\frac{G_{\sigma}(z-d,t)+G_{\sigma}(z+d,t)}{\sqrt{2}}\label{eq:Double-Well-Wavefunction}
\end{equation}
where $G_{\sigma}$ denotes the expanding Gaussian wave packet, 
\begin{equation}
G_{\sigma}(x,t)=\frac{\exp\left(-\frac{x^{2}}{4\sigma s_{t}}\right)}{\left(2\pi s_{t}^{2}\right)^{1/4}}\text{ with }s_{t}=\sigma+\frac{i\hbar t}{2m\sigma}.\label{eq:Expanding-Gaussian-Packet}
\end{equation}
In both cases, it is assumed that the confining wells are immediately
switched off at time $t=0$, letting the particles free thereafter.
In the case of the double well, it is assumed that the two wells are
well separated ($d\gg\sigma$), so that the free expansion leads to
an interference pattern in the density. One can check in Fig.~\ref{fig:Double-Well}
that this interference pattern is correctly reproduced by both the
Bohmian and stochastic trajectories.

\section{Arrival time distribution\label{sec:Arrival-Time-Distribution}}

\begin{figure*}
\includegraphics[height=5cm]{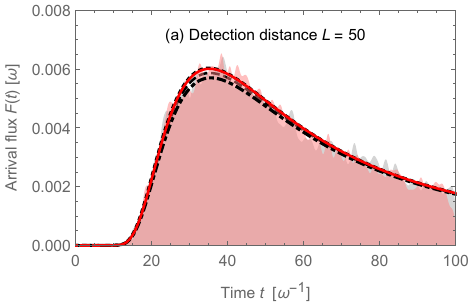}\hfill{}\includegraphics[height=5cm]{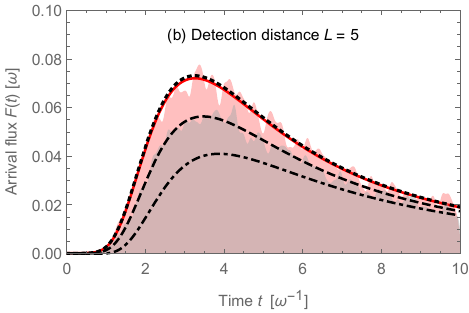}

\hfill{}\includegraphics[height=5cm]{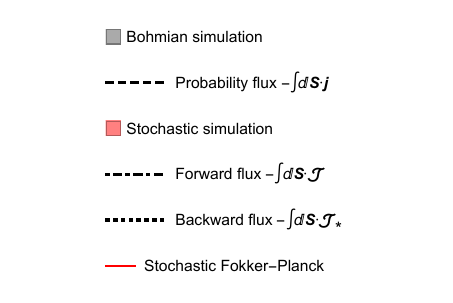}\hfill{}\includegraphics[height=5cm]{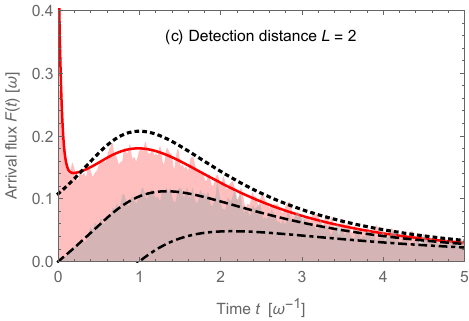}

\caption{\label{fig:Single-Well-Flux}Arrival flux of a cloud of particles
released from a single well of frequency $\omega=\hbar/2m\sigma^{2}$
and oscillator length $\sigma$, onto a detector at different distances
$L$ from the well. (a) $L=50\sigma$; (b) $L=5\sigma$; (c) $L=2\sigma$.
The Bohmian (grey fill) and stochastic (red fill) arrival fluxes are
obtained by sampling 32,000 trajectories propagated from Eqs.~(\ref{eq:Bohmian-Trajectory})
and (\ref{eq:Stochastic-Trajectory}) with $dt=1/(1600\omega)$.}
\end{figure*}
\begin{figure*}
\includegraphics[viewport=0bp 22bp 450bp 167bp,clip,width=16cm]{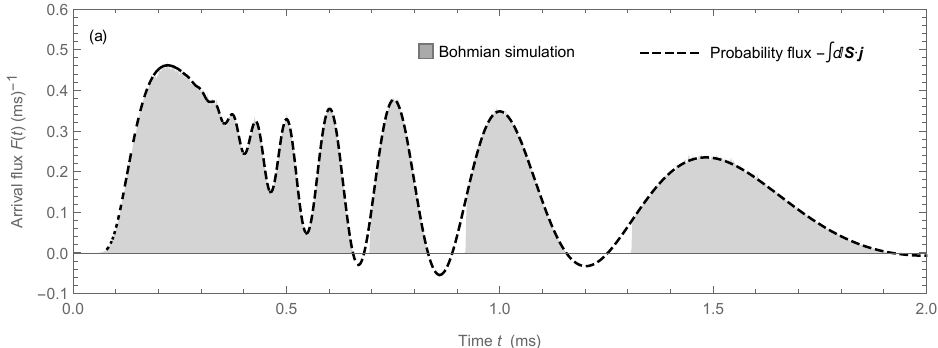}

\includegraphics[width=16cm]{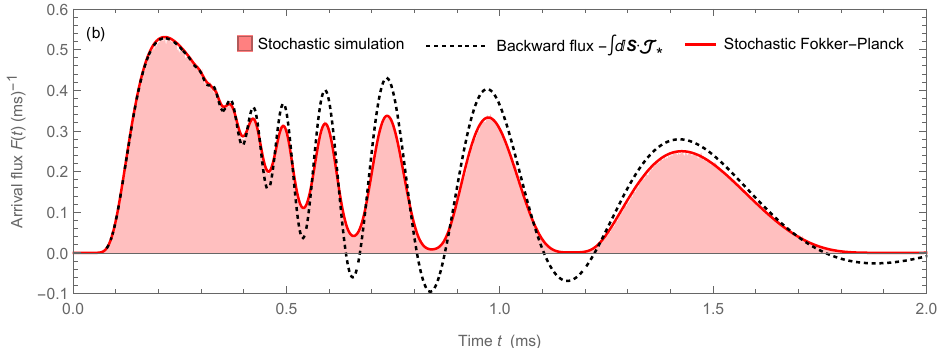}

\caption{\label{fig:Double-Well-Flux}Arrival flux of a cloud of particles
released from a double well, onto a detector at a distance $L=15\mu$m
from the centre of the two wells: (a) case of Bohmian trajectories
; (b) case of stochastic trajectories. The fluxes are obtained by
sampling $5\times10^{6}$ trajectories propagated from Eqs.~(\ref{eq:Bohmian-Trajectory})
and (\ref{eq:Stochastic-Trajectory}) with $dt=3.75\,\mu$s.}
\end{figure*}
Now let us consider the arrival time of the particle on a detector.
As mentioned earlier, the determination of arrival times is, in principle,
straightforward since the particle's possible trajectories are known.
Nevertheless, one is immediately faced with several important assumptions
about the detector that can affect the measured arrival times. Is
the detector localised around a single point, or a two-dimensional
plane? Is the particle ``destroyed'' by the detector when it is
detected (in the sense that it cannot be detected again)? Does the
detector affect the particle's motion? If so, does it simply select
trajectories guided by the wave function, or does it directly affect
the wave function itself?

In the following, it will be assumed that the detector is planar,
destroys the particle as soon as it is detected, but does not affect
its wave function (i.e., the wave function is assumed to remain the
same as in the absence of detector). Let us say that the detector
is placed at a certain distance $L$ below the source of the particle.
Then, according to our assumptions, the particle can only arrive from
above the detector plane (assuming that the detector plane is large
enough to prevent any trajectory from going around the detector and
hitting it from below). For a statistical distribution of trajectories,
the arrival time distribution is then simply proportional to the arrival
flux of trajectories hitting the detector plane from above. Note that
the motion in the three spatial directions are independent due to
the separability of the wave functions~Eqs.~(\ref{eq:Single-Well-Wavefunction})
and (\ref{eq:Double-Well-Wavefunction}), so that one can simply consider
the motion along the $z$ direction, as far as the arrival times on
the detector are concerned. 

In the case of Bohmian trajectories, it has been shown~\cite{Daumer1997}
that the flux of trajectories  through a plane is simply the flux
of the probability current $\bm{j}$ through that plane. When all
trajectories hit the detector plane from above, as in the case of
a particle released from a single harmonic well {[}see Fig.~\ref{fig:Single-Well}(b){]},
the arrival flux $F(t)$ is thus the flux $-\int_{P}d\bm{S}\cdot\bm{j}(\bm{x},t)$
of $\bm{j}$ through the detection plane $P$. Here, $d\bm{S}=dS\bm{n}$,
where $dS$ is the surface integration element and $\bm{n}$ the unit
vector orthogonal to the detection plane and pointing out from the
detecting side. Figure~\ref{fig:Single-Well-Flux} confirms that
the arrival flux of Bohmian trajectories numerically simulated from
Eq.~(\ref{eq:Bohmian-Trajectory}) (grey fill) coincides with the
 flux of the probability current $\bm{j}$ (dashed curve).  As discussed
in Ref.~\cite{AyatollahRafsanjani2023}, the situation is more complicated
in the case of the double well, because some trajectories may cross
the detection plane three times {[}see Fig.~\ref{fig:Double-Well}(b){]},
thus hitting once the detector plane from below (a situation known
as quantum reentry~\cite{Goussev2019,Goussev2020}). According to
the assumption that the particle is destroyed as soon as it first
hits the detector from above, the subsequent contributions from these
trajectories to the flux of $\bm{j}$ through the plane should be
discarded in the calculation of the arrival flux. This can only be
achieved through the simulation of many trajectories, as shown in
Fig.~\ref{fig:Double-Well-Flux}(a). One can see that the obtained
arrival flux (grey fill) is zero at, and slightly after, the arrival
times where the flux of $\bm{j}$ (dashed curve) is negative, because
the corresponding trajectories are blocked by the detector. Away from
these specific arrival times, the arrival flux is well reproduced
by the flux of $\bm{j}$.

In the case of stochastic trajectories, one could think by comparing
Eqs.~(\ref{eq:Bohmian-Trajectory}) and (\ref{eq:Stochastic-Trajectory})
that the arrival flux would be given by the forward current $\bm{\mathcal{J}}$.
However, that it is not the case. As shown in Fig.~\ref{fig:Single-Well-Flux},
the arrival flux of stochastic trajectories numerically simulated
from Eq.~(\ref{eq:Stochastic-Trajectory}) (red fill) is in fact
better approached by the flux of the backward current $\bm{\mathcal{J}}_{*}$
(dotted curve) than the forward current $\bm{\mathcal{J}}$ (dot-dashed
curve). This makes sense when one realises that the backward current
corresponds to the average current \emph{arriving} at a given point,
whereas the forward current corresponds to the average current \emph{departing}
from that point. Yet, the backward current only provides an approximation
of the arrival flux.

It is actually possible to determine the arrival flux exactly by considering
the density $\rho_{L}(\bm{x},t)$ of trajectories \emph{that do not
reach the detection plane}. That is because once the density of such
trajectories is known at a certain instant, one can calculate the
number of those first reaching the plane at the next instant. As shown
in Appendix~\ref{sec:Density-from-trajectories}, the density $\rho_{L}$
satisfies the following forward Fokker-Planck equation, 
\begin{equation}
\frac{\partial\rho_{L}}{\partial t}+\bm{\text{\ensuremath{\nabla}}}\cdot\left(\bm{b}\rho_{L}\right)-\frac{\hbar}{2m}\nabla^{2}\rho_{L}=0,\label{eq:Fokker-Planck-Equation}
\end{equation}
which is also known to be satisfied by the full density $\rho(\bm{x},t)$
of all possible trajectories~\cite{Nelson1966}. However, here it
is complemented by the following Dirichlet boundary condition at the
detection plane, $\rho_{L}(\bm{x},t)=0$ $\forall\bm{x}\in P$, which
effectively implements the restriction that the underlying trajectories
cannot reach the plane. It can be shown (see Appendix~\ref{sec:Density-from-trajectories})
that the first-arrival flux $F(t)$ at the plane (for trajectories
that do reach the plane) is then given by
\begin{equation}
F(t)=-\int_{P}d\bm{S}\cdot\frac{\hbar}{2m}\bm{\nabla}\rho_{L}(\bm{x},t).\label{eq:Stochastic-Flux}
\end{equation}
Figures~\ref{fig:Single-Well-Flux} and \ref{fig:Double-Well-Flux}(b)
show that the flux of Eq.~(\ref{eq:Stochastic-Flux}) obtained by
solving numerically the Fokker-Planck equation (\ref{eq:Fokker-Planck-Equation})
(red curve) agrees within the sampling errors with the one calculated
from the simulation of stochastic trajectories from Eq.~(\ref{eq:Stochastic-Trajectory})
(red fill). Unlike the case of Bohmian trajectories, the arrival time
distribution of stochastic trajectories can thus be obtained without
resorting to a sampling of trajectories.

\section{Experimental observation\label{sec:Experimental}}

Figures~\ref{fig:Single-Well-Flux} and \ref{fig:Double-Well-Flux}
clearly demonstrate that the arrival fluxes of Bohmian and stochastic
trajectories are in general different. However, they may be difficult
to distinguish experimentally. As found in Ref.~\cite{Nitta2008},
far from the source of particles, both the Bohmian and stochastic
fluxes are indistinguishable from the flux of the probability current
$\bm{j}$, as  seen in Fig.~\ref{fig:Single-Well-Flux}(a). They
become in principle distinguishable for detectors closer to the source,
as shown in Fig.~\ref{fig:Single-Well-Flux}(b), but remain largely
proportional to each other. Unless the initial number of particles
is precisely known and the detection is 100\% efficient, one could
only extract the arrival time distribution from the detector's counts,
which would not be conclusive. Much closer to the source, as in Fig.~\ref{fig:Single-Well-Flux}(c)
where the distance $L$ is only twice the trap width $\sigma$, the
arrival time distributions are predicted to be noticeably different.
However, besides the technical issues with implementing such a close
detector, the assumption that the detector does not alter the wave
function is questionable in this case.
\begin{figure*}
\includegraphics[width=16.7cm]{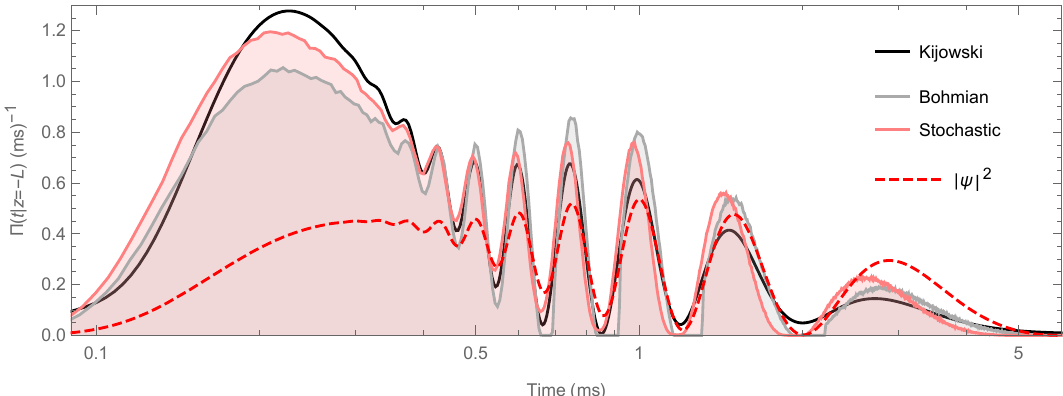}

\includegraphics[width=17cm]{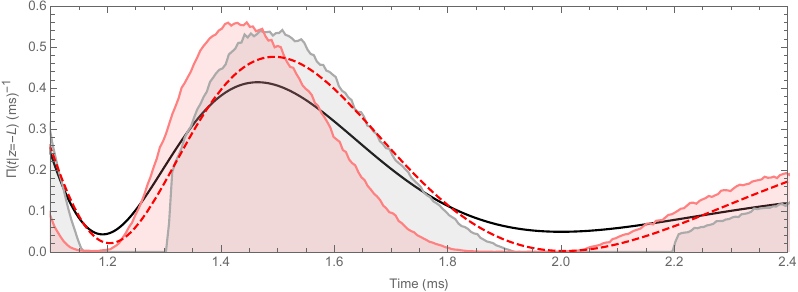}

\caption{\label{fig:Arrival-time-distribution}Top: Arrival time distribution
normalised over 6~ms, for different theories: the Kijowski arrival
time distribution (black) obtained in several theories of the quantum
arrival time~\cite{Aharonov1961,Allcock1969b,Kijowski1974,Grot1996,Delgado1997},
and the first-arrival time distributions obtained from Bohmian (grey)
and stochastic (red) trajectories. The dashed curve shows the normalised
distribution proportional to the wave function density $\vert\psi(\bm{x},t)\vert^{2}$
integrated on the detection plane; this distribution corresponds to
the multiple-arrival time distribution of stochastic trajectories,
as well as the as the distribution obtained in the quantum clock proposal~\cite{Roncallo2023}.
Bottom: Closeup of top panel around 2~ms, showing the time frames
where the Bohmian and stochastic distributions vanish.}
\end{figure*}

As advocated in Ref.~\cite{AyatollahRafsanjani2023}, the double-well
system with a detection plane perpendicular to the axis joining the
two wells is more promising. The authors propose to confine a cloud
of ultracold atoms in a double-well trap, release the trap and measure
the arrival times of the atoms on a detector. At ultracold temperature,
all atoms condense into the ground state of the trap. Each atom thus
constitutes a realisation of a single-particle time-of-flight experiment
for the same wave function. The best candidate for this purpose is
metastable helium-4, since it can be efficiently detected due to its
high internal energy~\cite{Vassen2012}. For easier comparison, the
same parameters as those chosen in Ref.~\cite{AyatollahRafsanjani2023}
have been taken in Figs.~\ref{fig:Double-Well}-\ref{fig:Double-Well-Flux}-\ref{fig:Arrival-time-distribution},
namely $m=4m_{u}$ (helium-4 mass, with $m_{u}$ the atomic mass unit),
$d=10\,\mu$m, $\sigma=0.5\,\mu$m. Note that the horizontal initial
velocity assumed in Ref.~\cite{AyatollahRafsanjani2023} to mimick
a double-slit experiment does not affect the vertical motion, and
is irrelevant in a double-well experiment where atoms are simply released
from their trap. For a detector at $L=15\mu$m from the centre of
the double well (thus at a distance 10$\sigma$ from the nearest well),
about 44\% of the released atoms reach the detector within 6~ms.
Their normalised arrival time distribution in that timeframe is shown
in Fig.~\ref{fig:Arrival-time-distribution}, for both the Bohmian
(grey) and stochastic trajectories (red). They are compared with the
\emph{Kijowski arrival time distribution} (black), which constitutes
an important reference since it can be derived from different approaches
to arrival time measurements, such as an axiomatic approach~\cite{Kijowski1974},
an absorption potential model~\cite{Allcock1969b}, a canonical quantisation
of the arrival time with the construction of a (generally non self-adjoint)
arrival-time operator~\cite{Aharonov1961,Grot1996}, or a self-adjoint
arrival-time operator that is not conjugate to the Hamiltonian~\cite{Delgado1997}.

One can see clear discrepancies between the three predicted distributions.
In particular trajectory-based interpretations of quantum theory both
predict time frames where there are no arrivals, in contrast with
the prediction of the Kijowski distribution. Observing a signal in
these timeframes would therefore invalidate these theories. To simulate
the statistical noise, the Bohmian and stochastic distributions are
calculated with the trajectories of $5\times10^{6}$ atoms, a number
typically achieved in experiments. One can see that this noise is
not an issue for distinguishing the curves. A major drawback is that
to experimentally rule out any of these curves, the experimental data
must be compared with a theoretical calculation. Thus, very precise
calibrations of all the parameters of the experiments, in particular
the position of the detector, trap frequency, and time of release,
are necessary. Nevertheless, the experimental discrimination appears
to be feasible in principle.

It is important to note that the above conclusions are bound to the
assumptions made in Section~\ref{sec:Arrival-Time-Distribution}.
Let us briefly discuss what to expect if these assumptions are invalidated.

If one assumes that the detector does not necessarily detect the particle
on its first arrival, but has a certain probability distribution for
detecting the particle on its various positions as it goes through
the detector, the resulting arrival time distribution may be quite
different, especially for stochastic trajectories. Indeed, while Bohmian
trajectories are smooth and cross the detector plane at most three
times, stochastic trajectories can enter the detector many times if
the particle is not immediately destroyed. It was found numerically
in Ref.~\cite{Nitta2008} that taking into account these multiple
counts through the detector gives a flux that is proportional to the
density $\vert\psi(\bm{x},t)\vert^{2}$ on the detector plane. This
result may sound surprising, since the density does not have the units
of a flux, as it was stressed earlier. The proportionality coefficient
must therefore have units of velocity. But what could be that constant
velocity? Although the work of Ref.~\cite{Nitta2008} did not address
this question, it is in fact possible to express analytically the
flux $F(t)$ of all stochastic trajectories through the detector plane
(see Appendix~\ref{sec:Flux-from-all-trajectories}). It turns out
that this flux is indeed proportional to the density, but the proportionality
coefficient is formally infinite: 
\begin{equation}
F(t)=\lim_{dt\to0}2\sqrt{\frac{\hbar}{2m\,\pi dt}}\int_{P}dS\vert\psi(\bm{x},t)\vert^{2}.\label{eq:multiple-arrival-flux}
\end{equation}
This means that the stochastic trajectories cross the detector plane
so many times that it results in an infinite flux. Physically, however,
the proportionality coefficient should be finite due to the limited
temporal resolution of the detector. Therefore, a detector detecting
many passages of a single particle is expected to yield a finite number
of counts proportional to the density. The first-arrival detection
and multiple-passage detection constitute two opposite limits. For
a detector detecting the particle with some delay probability distribution,
the measured arrival time distribution is expected to be somewhere
in-between these two limits. These limits are shown in Fig.~\ref{fig:Arrival-time-distribution}
by the solid red curve (immediate detection of the first arrival)
and the dashed red curve (multiple-passage detection). One can conclude
from that figure that even if the detector experiences delays and
multiple counts, the distribution resulting from stochastic trajectories
likely remains distinguishable from other predictions, although it
is more complicated to predict. Incidentally, let us remark that the
normalised distribution obtained from the flux of Eq.~(\ref{eq:multiple-arrival-flux})
coincides with the distribution obtained from the quantum clock proposal~\cite{Roncallo2023},
according to which time measurement is obtained from the entanglement
of the particle with a clock taken as a time reference.

Finally, there remains the question of whether the detector affects
the wave function. Some works~\cite{Werner1987,Tumulka2022a,Tumulka2022,Tumulka2024}
have proposed that the wave function is affected by the detector through
an absorbing boundary condition making the wave function proportional
to its gradient through the detector,
\begin{equation}
\bm{n}\cdot\bm{\nabla}\psi(\bm{x},t)=i\kappa\psi(\bm{x},t)\quad\forall\bm{x}\in P\label{eq:AbsorbingBoundary}
\end{equation}
where $\kappa$ is an inverse length characterising the detector.
It is clear that this condition makes the current $\bm{n}\cdot\rho\bm{\mathcal{V}}=\frac{\hbar}{m}\psi^{*}\bm{n}\cdot\bm{\nabla}\psi$
purely imaginary, i.e. $\bm{n}\cdot\bm{j}=\bm{n}\cdot\bm{\mathcal{J}}_{*}=\frac{\hbar}{m}\kappa\rho$.
This situation is similar to that of Fig.~\ref{fig:Single-Well-Flux}(a),
where $\bm{n}\cdot\bm{j}=\bm{n}\cdot\bm{\mathcal{J}}_{*}$ . It was
observed in that case that the stochastic and Bohmian trajectories
lead to indistinguishable arrival time distributions. It is therefore
likely that such a back effect of the detector on the wave function
would make it difficult to distinguish the predictions of stochastic
and Bohmian trajectories. However, it is presently unknown whether
the detector affects the wave function in this way.

\section{Conclusion}

\noindent This work shows that two trajectory-based interpretations
of quantum mechanics, the Bohmian and stochastic pilot wave theories,
do not in general yield the same arrival times. It appears that these
theories could be discriminated experimentally from other theories
of arrival time, as well as from each other, using ultra-cold atoms
released from a double well trap, as proposed in Ref.~\cite{AyatollahRafsanjani2023}.
Although questions remain regarding the role of the detection scheme,
it is an intriguing prospect that such experiments could shed some
light on the long-standing question of the existence and nature of
trajectories in quantum mechanics. %
\noindent\begin{minipage}[t]{1\columnwidth}%
\vspace{1cm}
\rule[0.5ex]{1\columnwidth}{1pt}%
\end{minipage}

\subsection*{Acknowledgments}

The author acknowledges support from the JSPS Kakenhi grant No. JP23K03292.

\noindent {\small{}{} }\bibliographystyle{IEEEtran2}
\bibliography{paper42}

\noindent {\small{} }{\small\par}

\clearpage{}

\begin{strip}

\renewcommand{\theequation}{\thesection.\arabic{equation}}
\setcounter{section}{0}
\setcounter{equation}{0}
\setcounter{figure}{0}\vspace{5mm}
\textsf{\textbf{\LARGE{}Appendix}}{\LARGE\par}

\vspace{5mm}
\end{strip}

\section{Stochastic motion\label{sec:Stochastic-motion}}

Let us consider a one-dimensional stochastic motion with forward drift
$b(x,t)$ and diffusion coefficient $\mathcal{D}=\frac{\hbar}{2m}$.
Namely, the position $x^{\prime}$ of the particle at time $t^{\prime}=t+dt$
is obtained from its position $x$ at time $t$ by the relation:
\begin{equation}
x^{\prime}=x+b(x,t)dt+\xi\label{eq:stochastic-process}
\end{equation}
where $\xi$ is a random number with average zero and variance $2\mathcal{D}dt$.
A realisation of such motion is shown in Fig.~\ref{fig:Example-of-stochastic}\@.

\section{Temporal distribution $F$\label{sec:Temporal-distribution}}

Let us define $F(x,t|x_{0},t_{0})dt$ as the probability for first
reaching $x$ between $t$ and $t+dt$ starting from $x_{0}$ at time
$t_{0}$. « First » means that $x$ has not been crossed along the
trajectory: the particle reaches $x$ for the first time between $t$
and $t+dt$ (see the green section of the curve in Fig.~\ref{fig:Example-of-stochastic}).

The probability for reaching $x$ between time $t_{0}$ and $t$ starting
from $x_{0}$ is therefore
\begin{equation}
0\le\int_{t_{0}}^{t}F(x,\tau|x_{0},t_{0})d\tau\le1\label{eq:probability-for-reaching-x}
\end{equation}

\section{Spatial distribution $R$\label{sec:Spatial-distribution}}

Let us now define $R(x,t|x_{0},t_{0})dx$ as the probability for reaching
$t$ between $x$ and $x+dx$, starting from $x_{0}$ at time $t_{0}$
(see the whole curve in Fig.~\ref{fig:Example-of-stochastic}). 

\subsection{Basic properties}

Since the particle must be somewhere at any time $t$, one must have

\begin{equation}
\int_{-\infty}^{\infty}dx\,R(x,t|x_{0},t_{0})=1\label{eq:R-normalisation}
\end{equation}
Moreover, by summing the probabilities for all possible positions
at an intermediate time $\tau$, one obtains the following chain rule:
\begin{equation}
\int_{-\infty}^{\infty}dy\,R(x,t|y,\tau)R(y,\tau|x_{0},t_{0})=R(x,t|x_{0},t_{0})\label{eq:R-chain-rule}
\end{equation}

\subsection{Fokker-Planck equation}

The probability distribution $R$ can be shown to satisfy the forward
Fokker-Planck equation~\cite{Hanggi1975}:
\begin{equation}
\frac{\partial R}{\partial t}+\frac{\partial}{\partial x}\left(bR\right)-\mathcal{D}\frac{\partial^{2}}{\partial x^{2}}R=0\label{eq:Fokker-Planck-equation-R}
\end{equation}
with initial condition $R(x,t_{0}\vert x_{0},t_{0})=\delta(x-x_{0})$.
Thus for any density $\rho(x,t)$ satisfying the above Fokker-Planck
equation with initial condition $\rho(x,t_{0})=\rho_{0}(x)$, one
can write $\rho(x,t)=\int dx_{0}R(x,t\vert x_{0},t_{0})\rho_{0}(x_{0})$.
For this reason, $R(x,t|x_{0},t_{0})$ may be regarded as the propagator
of the Fokker-Planck equation.

\subsection{Relation between $F$ and $R$}

\begin{figure}
\includegraphics[width=8cm]{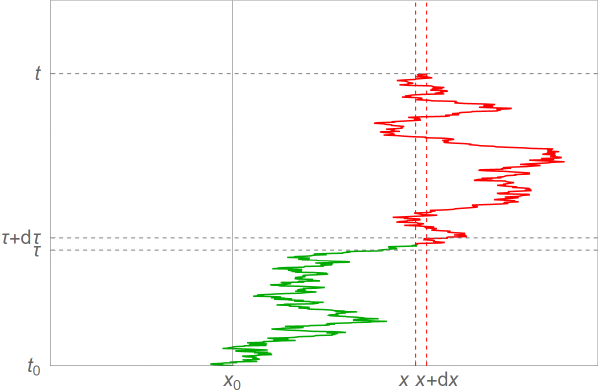}

\caption{\label{fig:Example-of-stochastic}Example of stochastic trajectory
going from $(x_{0},t_{0})$ and reaching $t$ between $x$ and $x+dx$.
It first reaches $x$ between time $\tau$ and $\tau+dt$, and then
crosses $x$ again several times between time $\tau+dt$ and $t$.
The ensemble of all possible such trajectories determines the spatial
probability distribution $R(x,t\vert x_{0},t_{0})$. The ensemble
of all possible trajectories going from $(x_{0},t_{0})$ and reaching
$x$ for the first time between $\tau$ and $\tau+dt$ (green section
of the curve) determines the temporal probability distribution $F(x,\tau\vert x_{0},t_{0})$.}
\end{figure}

One can find a relation between $F$ and $R$ by summing the probabilities
for all possible times at which the particle first reaches $x$ before
eventually reaching $x$ again at the final time $t$:
\begin{equation}
\int_{t_{0}}^{t}d\tau R(x,t|x,\tau)F(x,\tau|x_{0},t_{0})=R(x,t|x_{0},t_{0})\label{eq:Relation-between-F-and-R}
\end{equation}
This relation is illustrated in Fig.~\ref{fig:Example-of-stochastic}.

\section{Density from trajectories not reaching $y$\label{sec:Density-from-trajectories}}

The probability density for reaching $(x,t)$ starting from $(x_{0},t_{0})$
without having crossed $y$ at any time $\tau\in[t_{0},t]$ is given
by
\begin{equation}
R_{y}(x,t\vert x_{0},t_{0})\equiv R(x,t|x_{0},t_{0})-\tilde{R}(x,t\vert y|x_{0},t_{0})\label{eq:Definition-of-Ry}
\end{equation}
where 
\begin{equation}
\tilde{R}(x,t|y|x_{0},t_{0})=\int_{t_{0}}^{t}d\tau R(x,t|y,\tau)F(y,\tau|x_{0},t_{0})\label{eq:R-tilde}
\end{equation}
is the probability for reaching $(x,t)$ from $(x_{0},t_{0})$ and
having crossed $y$ at least once at some intermediate time $\tau$.
This implies that $R_{y}(y,t|x_{0},t_{0})=0$, owing to Eq.~(\ref{eq:Relation-between-F-and-R}).

Let us now calculate the following derivatives:
\begin{align}
\frac{\partial}{\partial t}\tilde{R}= & \underbrace{R(x,t|y,t)}_{\delta(x-y)}F(y,t|x_{0},t_{0})\label{eq:Derivative1}\\
 & +\int_{t_{0}}^{t}d\tau\frac{\partial R(x,t|y,\tau)}{\partial t}F(y,\tau|x_{0},t_{0})\nonumber \\
\frac{\partial}{\partial x}\left(b\tilde{R}\right)= & \int_{t_{0}}^{t}d\tau\frac{\partial}{\partial x}\left(bR(x,t|y,\tau)\right)F(y,\tau|x_{0},t_{0})\label{eq:Derivative2}\\
\frac{\partial^{2}}{\partial x^{2}}\left(\tilde{R}\right)= & \int_{t_{0}}^{t}d\tau\frac{\partial^{2}}{\partial x^{2}}\left(R(x,t|y,\tau)\right)F(y,\tau|x_{0},t_{0})\label{eq:Derivative3}
\end{align}
By summing the above equations and using the fact that $R$ satisfies
the Fokker-Planck equation~(\ref{eq:Fokker-Planck-equation-R}),
one arrives at
\begin{equation}
\frac{\partial}{\partial t}\tilde{R}+\frac{\partial}{\partial x}\left(b\tilde{R}\right)-\mathcal{D}\frac{\partial^{2}}{\partial x^{2}}\left(\tilde{R}\right)=\delta(x-y)F(y,t|x_{0},t_{0})\label{eq:Fokker-Planck-equation-Rtilde}
\end{equation}

Therefore, $R_{y}$ satisfies the Fokker-Planck equation
\begin{equation}
\frac{\partial}{\partial t}R_{y}+\frac{\partial}{\partial x}\left(bR_{y}\right)-\mathcal{D}\frac{\partial^{2}}{\partial x^{2}}\left(R_{y}\right)=0\label{eq:Fokker-Planck-equation-Ry}
\end{equation}
on $]-\infty,y[$ and $]y,\infty[$, with a discontinuity of spatial
derivative at $x=y$ given by
\begin{equation}
\left[\frac{\partial}{\partial x}R_{y}\right]_{x\to y^{+}}-\left[\frac{\partial}{\partial x}R_{y}\right]_{x\to y^{-}}=\frac{F(y,t|x_{0},t_{0})}{\mathcal{D}}\label{eq:derivative-discontinuity}
\end{equation}

Let us suppose that the starting point $x_{0}$ is on the left side
of $y$. Since by construction the particle cannot cross $y$, $R_{y}(x,t\vert x_{0},t_{0})$
must be identically zero for $x\ge y$, and the first term in Eq.~(\ref{eq:derivative-discontinuity})
should vanish. Therefore, for an arbitrary initial density distribution
$\rho_{0}(x)$ of points $x<y$, the density $\rho_{L}(x,t)\equiv\int_{-\infty}^{y}dx_{0}R_{y}(x,t\vert x_{0},t_{0})\rho_{0}(x_{0})$
is also identically zero for $x\ge y$ and satisfies the Fokker-Planck
equation 
\begin{equation}
\frac{\partial}{\partial t}\rho_{L}+\frac{\partial}{\partial x}\left(b\rho_{L}\right)-\mathcal{D}\frac{\partial^{2}}{\partial x^{2}}\left(\rho_{L}\right)=0\label{eq:Fokker-Planck-equation-rho-L}
\end{equation}
with the initial condition $\rho_{L}(x<y,t)\equiv\rho_{0}(x)$ and
the boundary condition $\rho_{L}(y,t\ge t_{0})=0$. The density $\rho_{L}$
is the density resulting from trajectories not crossing $y$ from
the left, starting from an initial density $\rho_{0}$. The temporal
distribution $F_{L}(y,t)$ for trajectories first reaching $y$ from
the left at time $t$ is therefore $F_{L}(y,t)\equiv\int_{-\infty}^{y}dx_{0}F(y,t\vert x_{0},t_{0})\rho_{0}(x_{0})$.
From Eq.~(\ref{eq:derivative-discontinuity}), one obtains:
\begin{equation}
F_{L}(y,t)=-\mathcal{D}\left[\frac{\partial\rho_{L}(x,t)}{\partial x}\right]_{x\to y^{-}}\label{eq:temporal-distribution-F-L}
\end{equation}

The formulation can be generalised to a three-dimensional stochastic
motion. The density $\rho_{L}(\bm{x},t)$ of three-dimensional trajectories
first reaching a plane $P$ from a given side (say left), starting
from an initial density $\rho_{0}(\bm{x})$, satisfies the three-dimensional
Fokker-Planck equation
\begin{equation}
\frac{\partial\rho_{L}}{\partial t}+\bm{\text{\ensuremath{\nabla}}}\cdot\left(\bm{b}\rho_{L}\right)-\mathcal{D}\nabla^{2}\rho_{L}=0.\label{eq:3d-Fokker-Planck-equation}
\end{equation}
with the boundary condition $\rho_{L}(\bm{x},t)=0\quad\forall\bm{x}\in P$.
The flux of these trajectories through the plane is then given by
\begin{equation}
F(t)=-\int_{P}d\bm{S}\cdot\mathcal{D}\bm{\nabla}\rho_{L}(\bm{x},t).\label{eq:Flux-through-plane}
\end{equation}
where $d\bm{S}=dS\bm{n}$ is the elementary surface vector pointing
from the left side of the plane.

\section{Flux from all trajectories\label{sec:Flux-from-all-trajectories}}

Let us now go back to the one-dimensional motion and consider the
flux of all (unblocked) trajectories through a certain point $x_{0}$
between time $t$ and $t+dt$. The basic idea of the calculation is
as follows. At time $t$, the probability density resulting from all
unblocked trajectories is known to given by $\rho(x,t)$. For very
small $dt$, the contributions to the flux between $t$ and $t+dt$
are given by trajectories coming from the neighbourhood of $x_{0}$,
typically from the range $[x_{0}-b(x,t)dt-\sqrt{\mathcal{D}dt},x_{0}-b(x,t)+\sqrt{\mathcal{D}dt}]$
since the trajectories diffuse during a time $dt$ within a typical
distance $\sim2\sqrt{\mathcal{D}dt}$. The number $d\mathcal{N}$
of such trajectories is therefore $\sim\rho(x_{0},t)2\sqrt{\mathcal{D}dt}$,
which gives a flux $d\mathcal{N}/dt\sim\rho(x_{0},t)2\sqrt{\mathcal{D}/dt}$
that is proportional to the local density $\rho(x_{0},t)$ but diverging
as $dt^{-1/2}$.

Here is a more precise derivation. The total number $d\mathcal{N}$
of trajectories starting at time $t$ from the density $\rho(x,t)$
and crossing the point $x_{0}$ (either from left or right) before
time $t+dt$ is given by
\begin{equation}
d\mathcal{N}=\int_{-\infty}^{\infty}dx\;\rho(x,t)P_{x_{0}}(x^{\prime}\vert x)\label{eq:dN-definition}
\end{equation}
where $P_{x_{0}}(x^{\prime}\vert x)$ is the probability of crossing
$x_{0}$ when starting from $x$ and ending at $x^{\prime}$ given
by the stochastic process of Eq.~(\ref{eq:stochastic-process}).
If the starting point $x$ is smaller than $x_{0}$, it is the probability
that $x^{\prime}>x_{0}$ and if the starting point $x$ is larger
than $x_{0}$, it is the probability that $x^{\prime}<x_{0}$, namely,
\begin{equation}
P_{x_{0}}(x^{\prime}\vert x)=\begin{cases}
P(x^{\prime}>x_{0}) & \text{for }x<x_{0}\\
P(x^{\prime}<x_{0}) & \text{for }x>x_{0}
\end{cases}\label{eq:Px0}
\end{equation}
From Eq.~(\ref{eq:stochastic-process}), one has $x^{\prime}=x+b(x,t)dt+\xi$,
and since $\xi\sim\sqrt{\mathcal{D}dt}$, for sufficiently small $dt$
one can neglect the term $b(x,t)dt$ with respect to $\xi$ in the
calculation of the probabilities. This gives
\begin{equation}
P_{x_{0}}(x^{\prime}\vert x)=\begin{cases}
P(\xi>x_{0}-x) & \text{for }x_{0}-x>0\\
P(\xi<x_{0}-x) & \text{for }x_{0}-x<0
\end{cases}\label{eq:Px0-bis}
\end{equation}

Even if the probability distribution of $\xi$ is not normal at very
small time scale, for $dt$ larger than that time scale, it becomes
normal due to the central limit theorem. Thus one can write
\begin{align}
P_{x_{0}}(x^{\prime}\vert x) & =P(\xi>\left|x_{0}-x\right|)\label{eq:Px0-ter}\\
 & =\int_{\vert x_{0}-x\vert}^{\infty}\frac{1}{\sqrt{4\pi\mathcal{D}dt}}\exp\left(-\frac{\xi^{2}}{4\mathcal{D}dt}\right)d\xi\nonumber \\
 & =\frac{1}{2}\text{erfc}\left(\frac{\left|x_{0}-x\right|}{\sqrt{4\mathcal{D}dt}}\right)\nonumber 
\end{align}
Now one can make a Taylor expansion of $\rho(x,t)$ around $x_{0}$
by setting $x=x_{0}+\epsilon$ in Eq.~(\ref{eq:dN-definition}):
\begin{equation}
d\mathcal{N}=\int_{-\infty}^{\infty}d\epsilon\;\left[\rho(x_{0},t)+\epsilon\frac{\partial\rho}{dx}(x_{0},t)+O(\epsilon^{2})\right]P_{x_{0}}(x^{\prime}\vert x)\label{eq:Taylor-expansion-of-rho}
\end{equation}
Performing the integration over $\epsilon$ using the explicit expression
of $P_{x_{0}}$ given by Eq.~(\ref{eq:Px0-ter}), one arrives at:
\begin{equation}
d\mathcal{N}=\sqrt{\frac{4\mathcal{D}dt}{\pi}}\rho(x_{0},t)+0+O(dt^{3/2})\label{eq:dN}
\end{equation}
The total flux through $x_{0}$ at time $t$ is therefore:
\begin{equation}
\frac{d\mathcal{N}}{dt}=\sqrt{\frac{4\mathcal{D}}{\pi dt}}\rho(x_{0},t)+O(dt^{1/2})\label{eq:dNoverdt}
\end{equation}
which is divergent in the limit of small $dt$.

The result is generalised to three dimensions for the adirectional
flux through a surface $S$:
\begin{equation}
F=\sqrt{\frac{4\mathcal{D}}{\pi dt}}\int_{S}dS\rho(\bm{x},t)\label{eq:3d-flux}
\end{equation}

\end{document}